\documentclass[submission,copyright,creativecommons]{eptcs}
\providecommand{\event}{GCM 2023} % Name of the event you are submitting to

\usepackage{iftex}

\ifpdf
  \usepackage{underscore}         % Only needed if you use pdflatex.
  \usepackage[T1]{fontenc}        % Recommended with pdflatex
\else
  \usepackage{breakurl}           % Not needed if you use pdflatex only.
\fi

\title{Model-Driven Rapid Prototyping for Control Algorithms with the GIPS Framework (System Description)}

\author{Maximilian Kratz
\institute{Technical University of Darmstadt}
\institute{Real-Time Systems Lab\\
Darmstadt, Germany}
\email{maximilian.kratz@es.tu-darmstadt.de}
\and
Sebastian Ehmes
\institute{Technical University of Darmstadt}
\institute{Real-Time Systems Lab\\
Darmstadt, Germany}
\email{sebastian.ehmes@es.tu-darmstadt.de}
\and
Philipp Maximilian Menzel
\institute{Technical University of Darmstadt}
\institute{Darmstadt, Germany}
\email{philipp\_maximilian.menzel}
\email{@stud.tu-darmstadt.de}
\and
Andy Schürr
\institute{Technical University of Darmstadt}
\institute{Real-Time Systems Lab\\
Darmstadt, Germany}
\email{andy.schuerr@es.tu-darmstadt.de}
}

\def\titlerunning{Model-Driven Rapid Prototyping for Control Algorithms with the GIPS Framework}
\def\authorrunning{M. Kratz, S. Ehmes, P. Menzel \& A. Schürr}

\usepackage{cite}
\usepackage{amsmath,amsfonts}
\usepackage{algorithmic}
\usepackage{graphicx}
\usepackage{textcomp}
\usepackage{xcolor}

\usepackage[nolist]{acronym}
\usepackage[T1]{fontenc} % Fixes encoding error in bibliography

\usepackage{mathtools}  % Extended version of amsmath
\usepackage{amssymb}    % Extended character set

\usepackage{hyperref}
\usepackage[noabbrev,capitalise]{cleveref}

\Crefname{equation}{Eq.}{Eqs.}

\usepackage{siunitx}
\usepackage{tikz}
\newcommand*\circled[1]{\tikz[baseline=(char.base)]{
		\node[shape=circle,draw,inner sep=1pt,minimum height=1em] (char) {#1};}}

\usepackage[inline]{enumitem}

\Crefname{lstlisting}{Listing}{Listings}
\usepackage{subcaption}

\usepackage{etoolbox}

\newtoggle{toolnamerev}
\togglefalse{toolnamerev}
\newtoggle{langnamerev}
\newcommand{\acr}[1]{\acs{#1} (\acl{#1})}

\newcommand{\toolname}{\iftoggle{toolnamerev}{\acs{GIPS}}{\acr{GIPS}\toggletrue{toolnamerev}}}
\newcommand{\langname}{\iftoggle{langnamerev}{\acs{GIPSL}}{\acr{GIPSL}\toggletrue{langnamerev}}}

\newcommand{\etal}{et~al.~}

\makeatletter
\newcommand*{\org@overidelabel}{}
\let\org@overridelabel\@verridelabel
\@ifpackagelater{acronym}{2015/03/21}{% v1.41
  \renewcommand*{\@verridelabel}[1]{%
    \@bsphack
    \protected@write\@auxout{}{\string\AC@undonewlabel{#1@cref}}%
    \org@overridelabel{#1}%
    \@esphack
  }%
}{% older versions
  \renewcommand*{\@verridelabel}[1]{%
    \@bsphack
    \protected@write\@auxout{}{\string\undonewlabel{#1@cref}}%
    \org@overridelabel{#1}%
    \@esphack
  }%
}
\makeatother

\hypersetup{
    pdftitle={Model-Driven Rapid Prototyping for Control Algorithms with the GIPS Framework (System Description)},
    pdfsubject={GCM 2023 System Description Paper},
    pdfauthor={Maximilian Kratz, Sebastian Ehmes, Philipp Menzel, Andy Schürr},
    pdfkeywords={Model-Driven Software Engineering, Optimization, Graph Transformation, Peer-to-Peer Data Distribution, Rapid Prototyping, Topology Control},
    breaklinks=true
}

\begin{document}
\maketitle

% Acronyms
\begin{acronym}
    \acro{CNF}{Conjuctive Normal Form}
    \acro{CPU}[CPU]{Central Processing Unit}
    \acro{DSL}{Domain-Specific Language}
    \acrodef{EBNF}[EBNF]{Extended Backus-Naur Form}
    \acro{EMF}{Eclipse Modeling Framework}
    \acro{GIPS}{\textbf{G}raph-Based \textbf{I}LP \textbf{P}roblem \textbf{S}pecification Tool}
    \acro{GIPSL}{\textbf{G}raph-Based \textbf{I}LP \textbf{P}roblem \textbf{S}pecification \textbf{L}anguage}
    \acro{GT}[GT]{Graph Transformation}
    \acro{IGPM}{Incremental Graph Pattern Matching}
    \acro{ILP}[ILP]{Integer Linear Programming}
    \acro{LHS}{Left-Hand Side}
    \acro{LOC}{Lines Of Code}
    \acro{MDSE}[MDSE]{Model-Driven Software Engineering}
    \acro{MdVNE}[MdVNE]{Model-driven Virtual Network Embedding}
    \acro{MILP}{Mixed-Integer Linear Programming}
    \acro{ML}{Machine Learning}
    \acro{MT}{Model Transformation}
    \acro{NOC}{Number Of Characters}
    \acro{P2P}{Peer-To-Peer}
    \acro{PM}{Pattern Matching}
    \acro{RHS}{Right-Hand Side}
    \acro{SAT}{propositional \textbf{sat}isfiability problem}
    \acro{SMT}{Satisfiability Modulo Theories}
    \acro{VNR}[VNR]{Virtual Network Request}
	\newacroplural{VNR}[VNRs]{Virtual Network Requests}
\end{acronym}

\begin{abstract}
    Software engineers are faced with the challenge of creating control algorithms for increasingly complex dynamic systems, such as the management of communication network topologies.
    To support rapid prototyping for these increasingly complex software systems, we have created the GIPS (\textbf{G}raph-Based \textbf{I}LP \textbf{P}roblem \textbf{S}pecification) framework\footnote{GIPS - \url{https://gips.dev}} to derive some or even all of the building blocks of said systems, by using \ac{MDSE} approaches.
    Developers can use our high-level specification language \langname~to specify their desired model optimization as sets of constraints and objectives. 
    GIPS is able to derive executable (Java) software artifacts automatically that optimize a given input graph instance at runtime, according to the specification. 
    Said artifacts can then be used as system blocks of, e.g., topology control systems.
    In this paper, we present the maintenance of (centralized) tree-based peer-to-peer data distribution topologies as a possible application scenario for GIPS in the topology control domain. 
    The presented example is implemented using open-source software and its source code as well as an executable demonstrator in the form of a virtual machine is available on GitHub.
\end{abstract}

\acresetall

\section{Introduction}
\label{sec:introduction}

Software developers working on modern (self-adaptive) control algorithms must deal with the ever-growing complexity of these types of software systems \cite{Brun2009}.
In the software engineering domain, model-driven software engineering tools have long been established as a valid approach for tackling the challenges of growing software complexity.
Consequently, one could use the principles of \ac{MDSE} and apply them to the development of control algorithms in general.
Hence, we propose to support the process of developing control algorithms with our newly developed framework \toolname~\cite{gipsGCM}, which lends itself to a rapid prototyping approach following the idea of models@run.time\cite{modelsAtRuntime, Bencomo2019, Cheng2014}.
\toolname~embodies the \ac{MDSE} approach by automatically deriving (Java) runtime artifacts from a given high-level specification, using our \ac{DSL} \langname~that is able to specify \ac{GT} rules.
It uses typed and attributed graphs as input as well as output models and performs model transformations based on the formally founded and established \ac{GT} framework \cite{taentzer}.
In combination with the well-known \ac{ILP} optimization approach \cite{appliedMathematicalProgramming}, \toolname~can be used to obtain sets of model transformations that adhere to global and local constraints as well as deliver results that optimize a given cost function.
Since \toolname~operates on generic structures such as graphs and uses a generic optimization approach, it is not limited to a specific problem domain.
In this paper, we make use of the generic nature of \toolname~and present its application in the prototype development of an incremental algorithm that maintains (centralized) tree-based \ac{P2P} overlay networks for data distribution in a video streaming scenario, with the goal to reduce server load and increase robustness.
The idea is to build this algorithm as a control algorithm in the form of the well-known MAPE-K loop \cite{visionOfAutonomicComputing, Brun2009}, where most of the system building blocks are automatically generated from a high-level specification by \toolname.
In contrast to \cite{gipsGCM}, which showed a \toolname-based application in the domain of data center networks, this paper presents \toolname~as a means to facilitate model-driven rapid prototyping and gives some thoughts on how to use the results of the prototyping phase regarding (topology) control algorithms using an example.

The rest of the paper is structured as follows.
In \cref{prelim:ilp} and \cref{prelim:gt}, we discuss the necessary basics of \ac{ILP} and \ac{GT} to give some fundamental knowledge of the underlying technologies of \toolname.
\Cref{sec:framework} gives a rough overview of our new framework \toolname.
In \cref{sec:example-scenario}, we introduce the aforementioned example scenario of the \ac{P2P} document distribution for the streaming platform lectureStudio\footnote{lectureStudio - \url{https://www.lecturestudio.org}} and our MAPE-K-based solution to the problem.
This section is split into the problem description (\cref{subsec:problem-description}), a possible solution approach (\cref{subsec:solution}) together with a demonstration (\cref{subsec:demonstration}), and a discussion on further challenges in the development process (\cref{subsec:threats-to-validity}).
In \cref{sec:related-work}, we discuss related work.
Finally, \cref{sec:conclusion-future-work} sums up our contribution and gives possible future enhancements.

% Artifact/Demo URL
The example implementation and the demonstrator in the form of a virtual machine are publicly available on GitHub\footnote{GIPS GCM 2023 Artifact VM - \url{https://github.com/Echtzeitsysteme/gips-gcm-2023-artifact-vm}}.

\section{Preliminaries}
\label{sec:preliminaries}

\subsection{\acl{ILP}}
\label{prelim:ilp}

\acl{ILP} (\acs{ILP}) is an optimization approach that can be used to find the minimum (maximum) of an objective function $F: \mathbb{Z}^n\rightarrow\mathbb{R}$ by solving for an integer target vector $\vec{x} \in \mathbb{Z}^n$ while adhering to some constraints $f_j(\vec{x}) \leq 0\;(j=1,\ldots,m)$ \cite{appliedMathematicalProgramming, luenbergerLinear2016, bazaraa2011linear}.
\Cref{eq:ilp-general} shows its canonical form, where\enspace$\vec{b} \in \mathbb{R}^m$ and $\vec{c} \in \mathbb{R}^n$\enspace are vectors,\enspace$A \in \mathbb{R}^{m \times n}$\enspace is a coefficient matrix, and\enspace$\vec{x} \in \mathbb{Z}^n$\enspace is the solution as a vector.
\begin{equation}
    \text{minimize} \enspace \vec{c}^T\vec{x} \enspace \text{s.\,t.} \enspace A \vec{x} \leq \vec{b}, \enspace \vec{x} \geq 0, \enspace \text{and} \enspace \vec{x} \in \mathbb{Z}^n\label{eq:ilp-general}
\end{equation}
The goal of the optimization (minimization or maximization) can be converted to the respective counterpart if the objective function is multiplied by the factor $-1$.
If all entries of the vector\enspace$\vec{x}$\enspace have to be either $0$ or $1$, i.e., they are Boolean variables, the \ac{ILP} problem is called a bivalent linear problem \cite{appliedMathematicalProgramming}.
Tomaszek \etal\cite{tomaszek2021VneEnsuringCorrectness} showed that the network embedding problem can be encoded as a bivalent linear problem.
To achieve this problem formulation, we have to model the problem as a set of unknown integer variables, linear constraints, and a linear target function.
The problem can afterwards be solved as an integer linear problem using solvers like Gurobi\footnote{Gurobi Optimizer - \url{https://www.gurobi.com/solutions/gurobi-optimizer/}} (commercial) or GLPK\footnote{GNU Linear Programming Kit - \url{https://www.gnu.org/software/glpk/}} (free and open-source software).

\subsection{Graph Transformation} \label{prelim:gt}

\toolname~operates on models based on typed and attributed graphs, where objects correspond to typed nodes and references between objects correspond to typed edges.
For this reason, we can make use of \acp{GT}, which is a formal framework that provides a rule-based way to define model transformations on graphs.
\ac{GT} rules consist of a \ac{LHS} and a \ac{RHS}.
Both, the \ac{LHS} and the \ac{RHS} are graph patterns that describe together which structures must be present, absent, created, or deleted in a given target graph.
The \ac{LHS} defines which graph structures have to be present in a target graph before a \ac{GT} rule can be applied.
The \ac{RHS}, on the other hand, defines which graph structures must be present in the target graph after the rule has been applied.
Unsurprisingly, \ac{GT} heavily depends on graph \ac{PM}, which is used to find subgraphs in a given target graph that match a given graph pattern, e.g., the \ac{LHS} of a \ac{GT} rule.
Such a subgraph is called a match and consists of graph pattern nodes successfully mapped to a set of target graph nodes.
\toolname~relies on \ac{IGPM}, a commonly used \ac{PM} approach, which keeps track of individual model changes, to update sets of appearing or vanishing matches incrementally. 
In \toolname, we make use of the HiPE\footnote{HiPE - \url{https://github.com/HiPE-DevOps/HiPE-Updatesite}} pattern matching engine, which is based on a massively parallelized variant of Forgy's Rete-approach \cite{PatternMatchReteNetwork}.

\section{The GIPS Framework} % Hard-coded framework name instead of macro to remove PDF token warning
\label{sec:framework}

\begin{figure*}[hbt]
\centering
    \includegraphics[width=1.0\linewidth]{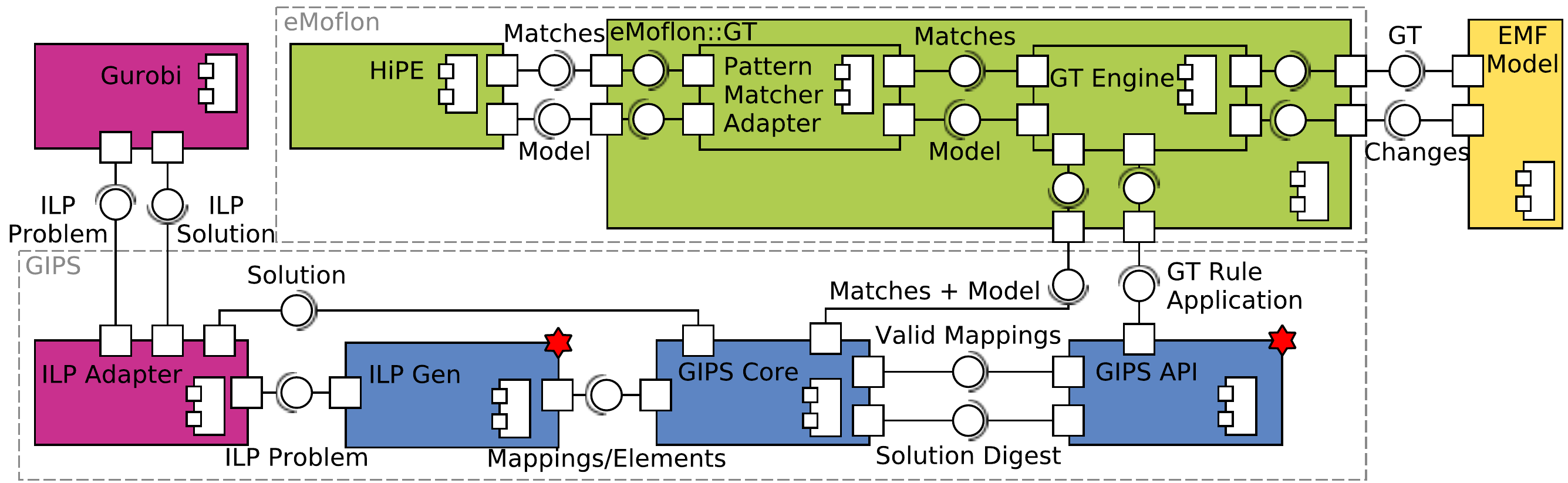}
    \caption{Component diagram of the \toolname~framework (red star = generated during build time) \cite{gipsGCM}.}
    \label{figure:gips-components}
\end{figure*}

The two driving motivations behind \toolname~are:
\begin{enumerate*}[label=(\arabic*)]
    \item Reducing the amount of effort (i.e., code) that has to be put into the creation of complex prototypes of graph optimization tools and
    \item a reduction of programming errors that will inevitably occur in sufficiently complex projects, regardless of the programmer's skills.
\end{enumerate*}
For this purpose, we have developed \langname~that enables expressing \ac{ILP} constraints in a UML/OCL-like fashion and which extends the eMoflon::IBeX-GT language\footnote{\toolname~uses eMoflon::IBeX as a \ac{GT} engine - \url{https://emoflon.org/ibex}} that provides the ability to specify \ac{GT} rules and graph patterns. 
As a result, we gain a level of integration that allows constraint definitions to access and, thus, make use of graph nodes of \ac{GT} rules or graph patterns.
In essence, an output model of \toolname~is shaped by a set of \ac{GT} rule applications, which describe local modifications to a given model and are only executed if a rule’s \ac{LHS} holds beforehand.
A subset of valid matches of a specific \ac{GT} rule’s \ac{LHS} will be determined by the \ac{ILP} solver.
Said matches are only valid if they adhere to all constraints defined in the corresponding \langname~specifications.
By imposing \ac{ILP} constraints that interweave a set of \ac{GT} rule applications, we can enrich the expressiveness of their respective \acp{LHS}, which usually only allows for local first-order logical conditions, due to missing knowledge of other possible rule applications.
Since graph pattern matches already fulfill certain localized structural constraints as defined by their corresponding graph patterns, we do not need to encode those local constraints in the \ac{ILP}.
Thus, building an \ac{ILP} problem from a limited set of graph pattern matches (i.e., tuples of graph nodes) instead of the whole graph can in most cases reduce the search space of the \ac{ILP} problem significantly, and, therefore, increase the performance of the \ac{ILP} optimization step.

The \toolname~framework itself consists of four components, which are shown in \cref{figure:gips-components}.
During build time, the \ac{ILP} generator component \texttt{ILP Gen} and the user API \texttt{GIPS API} are generated according to given \langname~specifications (as indicated with the red star symbol in \cref{figure:gips-components}).
The \texttt{GIPS Core} component interfaces with the incremental \ac{GT} tool eMoflon to get access to matches and the underlying model data.
The gathered set of matches and the model data are used by the generated \ac{ILP} generator component \texttt{ILP Gen} to construct a new \ac{ILP} problem.
Said \ac{ILP} formulation is then used by the \texttt{ILP Adapter} to connect to the \ac{ILP} solver to supply it with the problem and receive a valid solution (i.e., one that satisfies all constraints) if one exists.
Finally, \ac{GT} rules can be applied to matches by the use of the generated API \texttt{GIPS API} based on the solution.

\begin{figure*}[hbt]
\centering
    \includegraphics[width=1.0\linewidth]{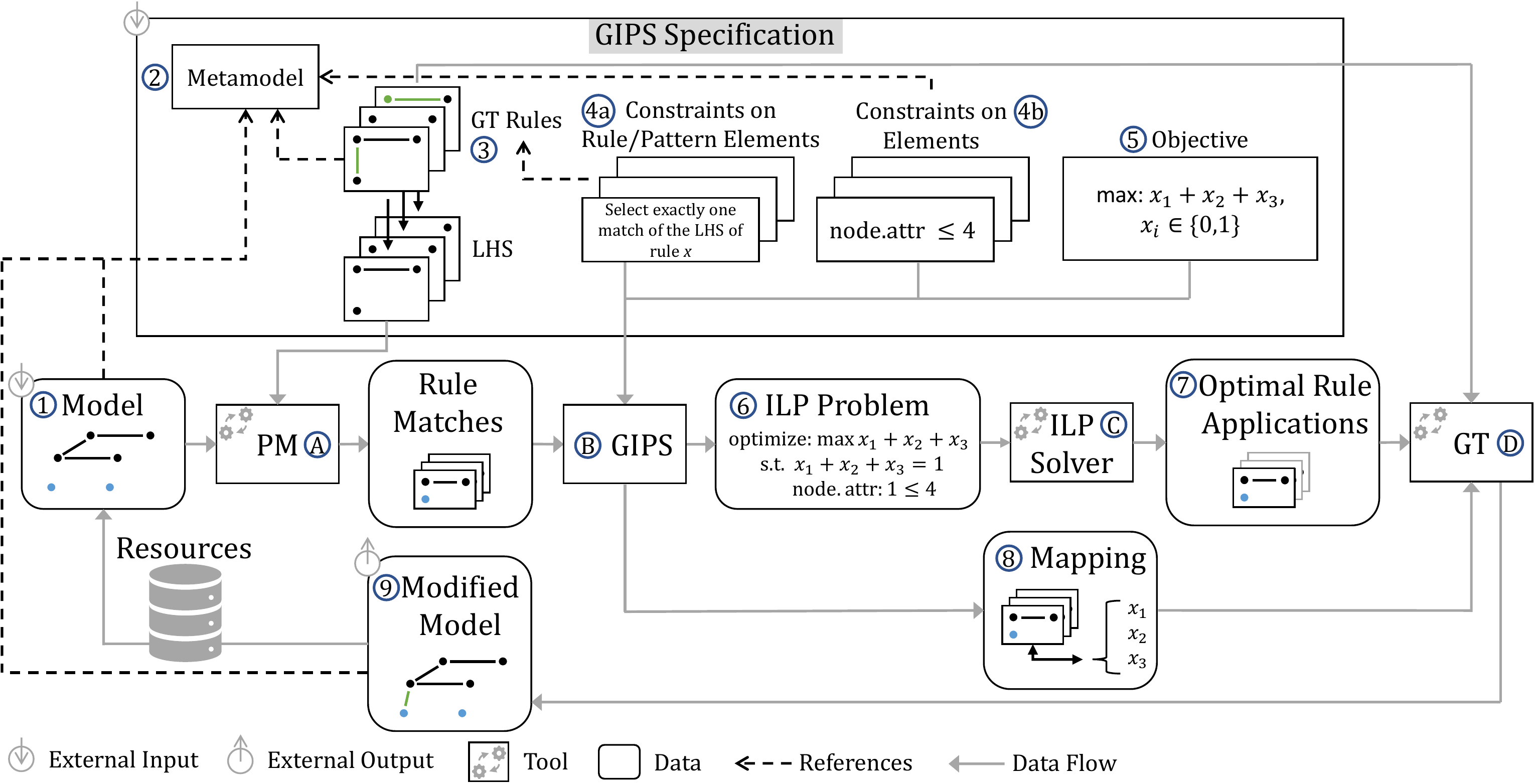}
    \caption{Extended workflow diagram of the \toolname~framework \cite{gipsGCM}.}
    \label{figure:gips-workflow}
\end{figure*}

In the following, we will roughly outline how \toolname~works, using \cref{figure:gips-workflow} as an illustration of an exemplary \toolname~workflow.
Let us assume three input artifacts, namely a metamodel \hyperref[figure:gips-workflow]{\circled{2}}, a corresponding graph-based model \hyperref[figure:gips-workflow]{\circled{1}} that is to be altered and the \langname~specification, which contains a set of \ac{GT} rules \hyperref[figure:gips-workflow]{\circled{3}}, a set of rule and model constraints \hyperref[figure:gips-workflow]{\circled{4}} as well as an objective function \hyperref[figure:gips-workflow]{\circled{5}}.
The exemplary constraint in \hyperref[figure:gips-workflow]{\circled{\small4a}} ensures that in the resulting set of \ac{GT} rule applications, the rule $x$ must be applied exactly once.
Furthermore, another exemplary constraint (see \hyperref[figure:gips-workflow]{\circled{\small4b}}) ensures that an attribute of a model element has to be smaller or equal to $4$.
The exemplary objective (see \hyperref[figure:gips-workflow]{\circled{5}}) consists of a linear function of \ac{ILP} decision variables ($x_i \in \{0,1\}$) that is to be maximized.
As shown in \cref{figure:gips-workflow}, the \toolname~process starts with requesting matches from an external incremental pattern matcher \hyperref[figure:gips-workflow]{\circled{A}}.
As a result, \toolname~has all valid rule matches and, thus, all locations where a \ac{GT} rule can be applied.
The matches alongside constraints and the objective are used by \toolname~\hyperref[figure:gips-workflow]{\circled{B}} to generate an \ac{ILP} problem \hyperref[figure:gips-workflow]{\circled{6}}, where each variable corresponds to a match.
In this particular \ac{ILP} problem example (see \hyperref[figure:gips-workflow]{\circled{6}}), the objective function is taken from \hyperref[figure:gips-workflow]{\circled{5}}, the first constraint is the result of \hyperref[figure:gips-workflow]{\circled{\small4a}}, and the second constraint is the result of \hyperref[figure:gips-workflow]{\circled{\small4b}}.
Additionally, mappings are created that encode which variable corresponds to which match \hyperref[figure:gips-workflow]{\circled{8}}.
As previously mentioned, \toolname~makes it possible to subject \ac{GT} rules to \ac{ILP} constraints, which limit rule applications \hyperref[figure:gips-workflow]{\circled{\small4a}}, e.g., to prevent rules from invalidating matches of other rules \acp{LHS}.
Besides that, it is also possible to impose \ac{ILP} constraints onto the model \hyperref[figure:gips-workflow]{\circled{\small4b}}, e.g., to enforce certain (aggregated) attribute values or to limit the total number of edges connected to a node.
An external \ac{ILP} solver \hyperref[figure:gips-workflow]{\circled{C}} will then calculate an optimal solution w.r.t.~the given objective function, ensuring all given constraints.
Finally, a solution to a given problem implicitly contains valid rule applications (matches) encoded as a set of non-zero binary variables \hyperref[figure:gips-workflow]{\circled{7}}.
This means that the selected rule applications are in arbitrary order.
Using these resulting binary variables in combination with the aforementioned mappings to their corresponding matches and \ac{GT} rules, an external \ac{GT} engine \hyperref[figure:gips-workflow]{\circled{D}} performs the graph modifications, which results in a modified model \hyperref[figure:gips-workflow]{\circled{9}}.

\section{Adapting P2P Overlay Networks for Data Distribution}
\label{sec:example-scenario}

In the following, we present our example scenario centered around (centralized) \ac{P2P} overlay network maintenance, to show that a model-driven software engineering approach, as implemented by \toolname, facilitates rapid prototyping of control algorithms.
To this end, we demonstrate that we can use \toolname~to specify and generate most components of a generic MAPE-K loop \cite{visionOfAutonomicComputing, Brun2009} (see \cref{figure:mape-k-gips}), which is a popular approach for control algorithms of the self-adaptive system's domain.
% MAPE-K
In short, a MAPE-K loop is a sequence of four stages \textit{Monitor}, \textit{Analyze}, \textit{Plan}, and \textit{Execute} over the \textit{Knowledge} base.
The latter maintains data on all relevant parts of the system that are needed by the MAPE stages.
The \textit{Monitor} collects data from the underlying system and the environment, whereas the \textit{Analyze} stage checks if an adaption of the system is required.
If this is the case, it triggers the \textit{Plan} stage to construct an adaption plan.
Lastly, the \textit{Execute} stage ensures the plan gets executed to adapt the system.

% Problem description
\subsection{Problem Description}
\label{subsec:problem-description}

\begin{figure}[ht]
\centering
    \includegraphics[width=0.66\linewidth]{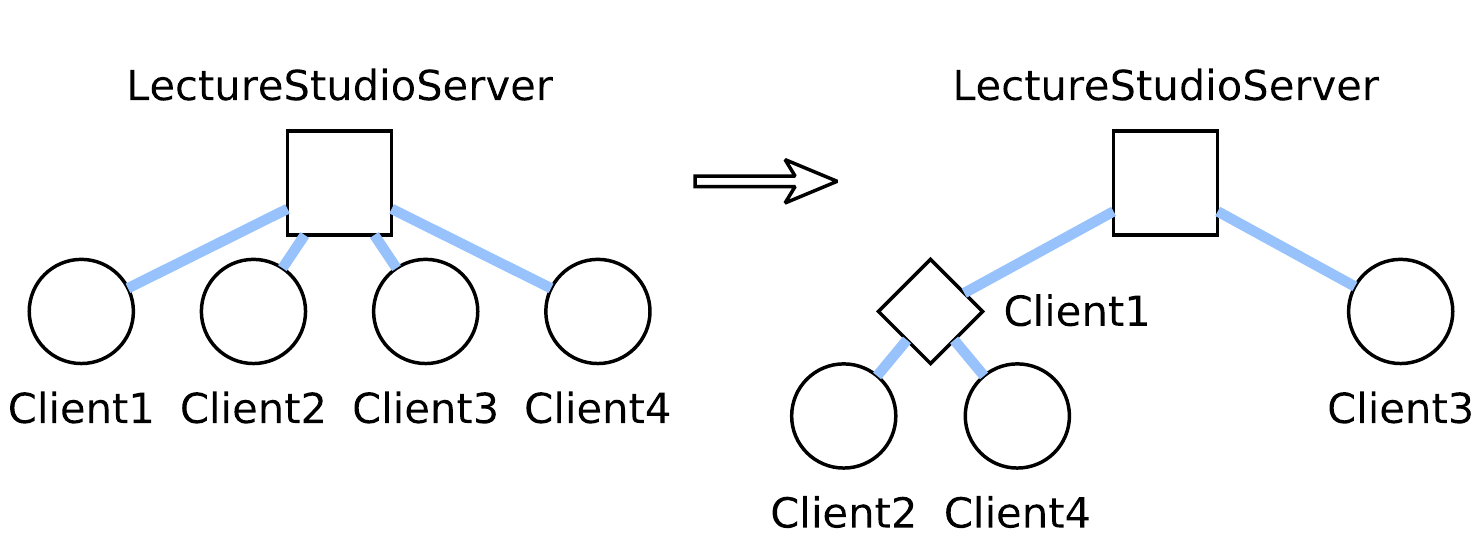}
    \caption{Small example network of a lectureStudio document distribution. The left one is purely centralized and the right one uses a peer-to-peer mechanism. The central server node (\texttt{LectureStudioServer}) is shown as a square, relay clients are shown as diamonds, and normal clients are shown as circles.}
    \label{figure:lectureStudio-example}
\end{figure}

In recent years, the need for online streaming of lectures and other events has grown considerably.
Therefore, the tool and streaming platform lectureStudio has been developed to support lecturers at the Technical University of Darmstadt.
One key aspect of this platform is the fact that it is not only able to stream continuous video, e.g., from the webcam of the lecturer or a shared screen but also their slides as a PDF file.
The difference to conventional video conference tools is the approach of distributing the used PDF slide set to all participants at the beginning of the lecture.
After the lecture has initially started, only small actions such as slide changes or annotation commands are sent to the clients.
By using this approach, the total used bandwidth of the streaming process can be reduced drastically.
However, as one might notice, the distribution of possibly large PDF files at the beginning of the lecture causes a huge spike in bandwidth demand for the lectureStudio server.
In case the document is large or the number of students participating in the stream is high, said operation can saturate the connection of the server for quite some time.
To circumvent this issue, the approach presented in this example calculates a tree-based \ac{P2P} overlay topology to allow client-to-client distribution of the PDF file.
By using such a \ac{P2P} approach, the central lectureStudio server only has to transfer the PDF document to a subset of all connected clients, which will transfer (relay) the file to the remaining clients afterwards.

\Cref{figure:lectureStudio-example} shows two examples of distribution overlay networks with square and diamond nodes indicating the (re-)distributors of PDF files.
The left one uses a classic purely centralized approach in which all clients download the PDF file directly from the server (\texttt{LectureStudioServer}).
The right network contains a relay client (\texttt{Client1}) that forwards the data to two other clients in order to free up some of the server's bandwidth.
Hence, our example implementation automatically derives a \ac{P2P} topology from a given topology and its changes over time, while optimizing a specific objective function to minimize the file distribution time for all participating clients.
Therefore, the central algorithm that calculates the \ac{P2P} overlay topology must determine
\begin{enumerate*}[label=(\arabic*)]
    \item which clients must become relay clients and
    \item how clients connect to other nodes
\end{enumerate*}
in order to optimize a specific objective function, in this case, to minimize the file distribution time for all clients.

% Solution
\subsection{Prototype Implementation}
\label{subsec:solution}

\begin{figure}[hbt]
\centering
    \includegraphics[width=0.75\linewidth]{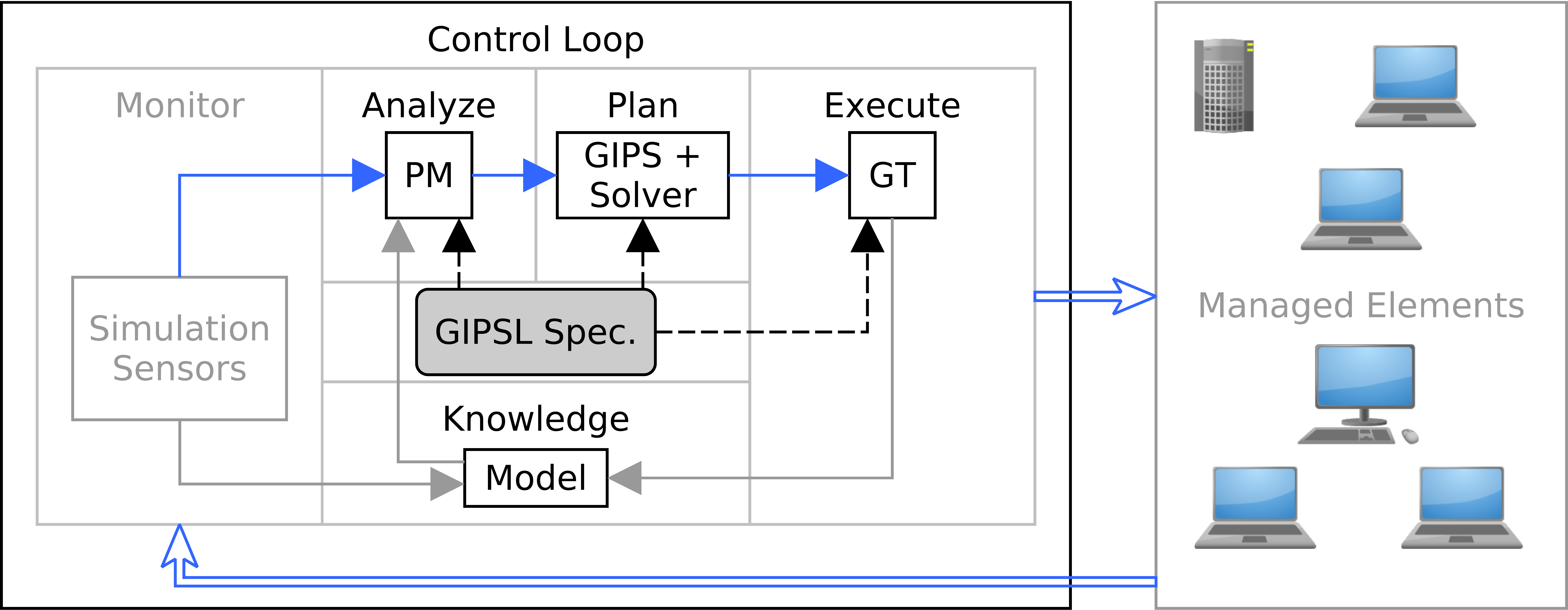}
    \caption{\toolname~as MAPE-K loop with the lectureStudio server (control loop) and the clients (managed elements). Blue arrows indicate data flow.}
    \label{figure:mape-k-gips}
\end{figure}

In this section, we present our artifact, which implements an algorithm that solves the challenge described in \cref{subsec:problem-description}.
Our algorithm is based on a MAPE-K loop as shown in \cref{figure:mape-k-gips}, in which the \ac{P2P} overlay network is incrementally adapted as a reaction to model changes.
As indicated in the figure, we used \toolname~to generate the building blocks \texttt{Analyze}, \texttt{Plan}, and \texttt{Execute} of the loop from a high-level \langname~specification, which makes \toolname~an integral part of the submitted artifact.
The role of the control loop is played by the lectureStudio server, which coordinates the behavior of all clients that are part of the managed elements.
Currently, the whole example and its implementation are only part of a simulation.
Hence, all managed elements (clients) only exist as objects in the model and there is no connection to an actual instance of the lectureStudio server application.

\begin{figure}[hbt]
\centering
    \includegraphics[width=0.75\linewidth]{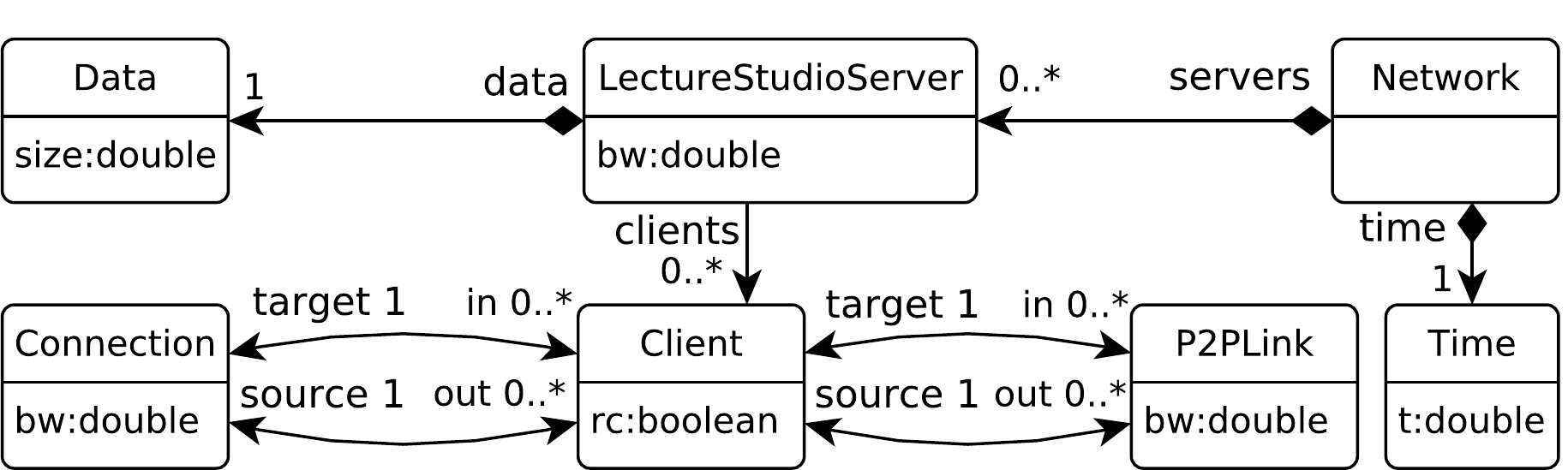}
    \caption{Simplified metamodel of the example's implementation.}
    \label{figure:metamodel}
\end{figure}

Our knowledge component is a graph-based model implemented using the \ac{EMF}\footnote{Eclipse Modeling Framework - \url{https://www.eclipse.dev/modeling/emf/}} (\hyperref[figure:gips-workflow]{\circled{1}} in \cref{figure:gips-workflow}).
This model corresponds to a metamodel (\hyperref[figure:gips-workflow]{\circled{2}} in \cref{figure:gips-workflow}), which defines the model’s node types, attributes, and edges, similar to a UML class diagram.
As shown in \cref{figure:metamodel}, our network metamodel (which is an example of \hyperref[figure:gips-workflow]{\circled{2}} in \cref{figure:gips-workflow}) consists of the following nodes\footnote{The complete (non-shortened) metamodel can be found in this repository: \url{https://github.com/Echtzeitsysteme/gips-gcm-2023-example}}:
\texttt{Network}, \texttt{LectureStudioServer}, \texttt{Client}, \texttt{Connection}, \texttt{P2PLink}, \texttt{Data}, and \texttt{Time}.
A node of type \texttt{Network} contains the central node of type \texttt{LectureStudioServer}, each containing a set of clients represented by nodes of type \texttt{Client}.
Nodes of type \texttt{Client} can either be normal clients or relay clients; in the latter case, the Boolean flag \texttt{rc} of the respective node is set to \texttt{true}.
Moreover, nodes of type \texttt{Connection} connect two nodes of type \texttt{Client} via \texttt{source} and \texttt{target} edges, and model the maximum possible bandwidth with the attribute \texttt{bw}.
(Virtual) \ac{P2P} connections are modeled by nodes of type \texttt{P2PLink}, which also connect two nodes of type \texttt{Client} with an attributed bandwidth \texttt{bw}.
A \texttt{LectureStudioServer} contains one node of type \texttt{Data}, which models the PDF file to be transferred when the lecture starts.
Finally, a node of type \texttt{Time} contains a global variable \texttt{t} that is used to calculate the actual number of time steps needed to distribute data to all clients in the simulation.
\Cref{figure:lectureStudio-example} is a simplified example of a network model that corresponds to the metamodel of \cref{figure:metamodel}.

% GIPSL specification
To generate the missing blocks of the MAPE-K loop, namely \textit{Analyze}, \textit{Plan}, and \textit{Execute} using \toolname~(see \cref{figure:mape-k-gips}), we need a \langname~specification that defines the behavior of said stages through \ac{GT} rules, patterns, \ac{ILP} constraints, and an objective function to optimize.

In general, the \textit{Analyze} stage of a MAPE-K loop identifies locations in the model that violate constraints and, in turn, it also detects locations that can be modified to reach a constraint-compliant state again.
Moreover, the \textit{Analyze} stage also detects locations that might be used to re-establish an optimal system state (according to the objective function), in case the system has lost this property due to external changes.
In our case, the outcome of the \textit{Analyze} stage is largely defined by the \acp{LHS} of the \ac{GT} rules as well as the patterns used in the \langname~specification (\hyperref[figure:gips-workflow]{\circled{3}} in \cref{figure:gips-workflow}).
Effectively, the output of this stage is a collection of matches (see the connection of \hyperref[figure:gips-workflow]{\circled{A}} and \hyperref[figure:gips-workflow]{\circled{B}} in \cref{figure:gips-workflow}) each describing a possible location in the input model for the application of a corresponding \ac{GT} rule, which could possibly repair a constraint violation or improve the objective function value.
For example, a specific rule can contain instructions to modify the model in such a way that a waiting \texttt{Client} will be connected to the \texttt{LectureStudioServer} instance, which works towards satisfying the constraint that requires each \texttt{Client} in the network being (indirectly) connected to the \texttt{LectureStudioServer} node.
During runtime, this will instruct the \ac{IGPM} (\hyperref[figure:gips-workflow]{\circled{A}} in \cref{figure:gips-workflow}) to search the input model for matches of this rule’s \ac{LHS}, which will each contain a \texttt{Client} and a \texttt{LectureStudioServer} node.

The \textit{Plan} stage uses the matches of the \textit{Analyze} stage along with the constraints (\hyperref[figure:gips-workflow]{\circled{\small4a}} and \hyperref[figure:gips-workflow]{\circled{\small4b}} in \cref{figure:gips-workflow}) and the objective function (\hyperref[figure:gips-workflow]{\circled{5}} in \cref{figure:gips-workflow}) of the \langname~specification to generate the actual \ac{ILP} problem (\hyperref[figure:gips-workflow]{\circled{6}} in \cref{figure:gips-workflow}) that will be solved with the help of an \ac{ILP} solver (\hyperref[figure:gips-workflow]{\circled{C}} in \cref{figure:gips-workflow}).
The overarching goal of the \textit{Plan} stage is to select the best valid rule applications (i.e., rule-match-pairs, \hyperref[figure:gips-workflow]{\circled{7}} in \cref{figure:gips-workflow}) from the \textit{Analyze} stage.
In our example, we used \langname~to define \(3\)\,\ac{GT} rules, \(8\)\,graph patterns, and \(9\)\,constraint specifications, to model the following behaviour:
For each newly appearing or vanishing client, our planner tries to select the subset of all possible rule applications that extend or repair the overlay network, such that it minimizes the average bandwidth of all used connections between all nodes while adhering to all specified constraints.
Thus, the output of this stage is a set of matches whose corresponding \ac{GT} rules must be applied in the \textit{Execute} stage.
As a side note, \toolname, in general, does not provide a specific order in which the resulting set of rule applications has to be executed.
Either users build their own execution logic using the \toolname~generated API, or the \langname~specification is defined in such a way that the execution order does not matter.
In our example implementation, we chose the latter option and specified the \ac{ILP} problem in such a manner that we may execute our set of valid rule applications in an arbitrary order.

% ILP example constraints
\begin{figure}[ht]
\centering
    \includegraphics[width=0.55\linewidth]{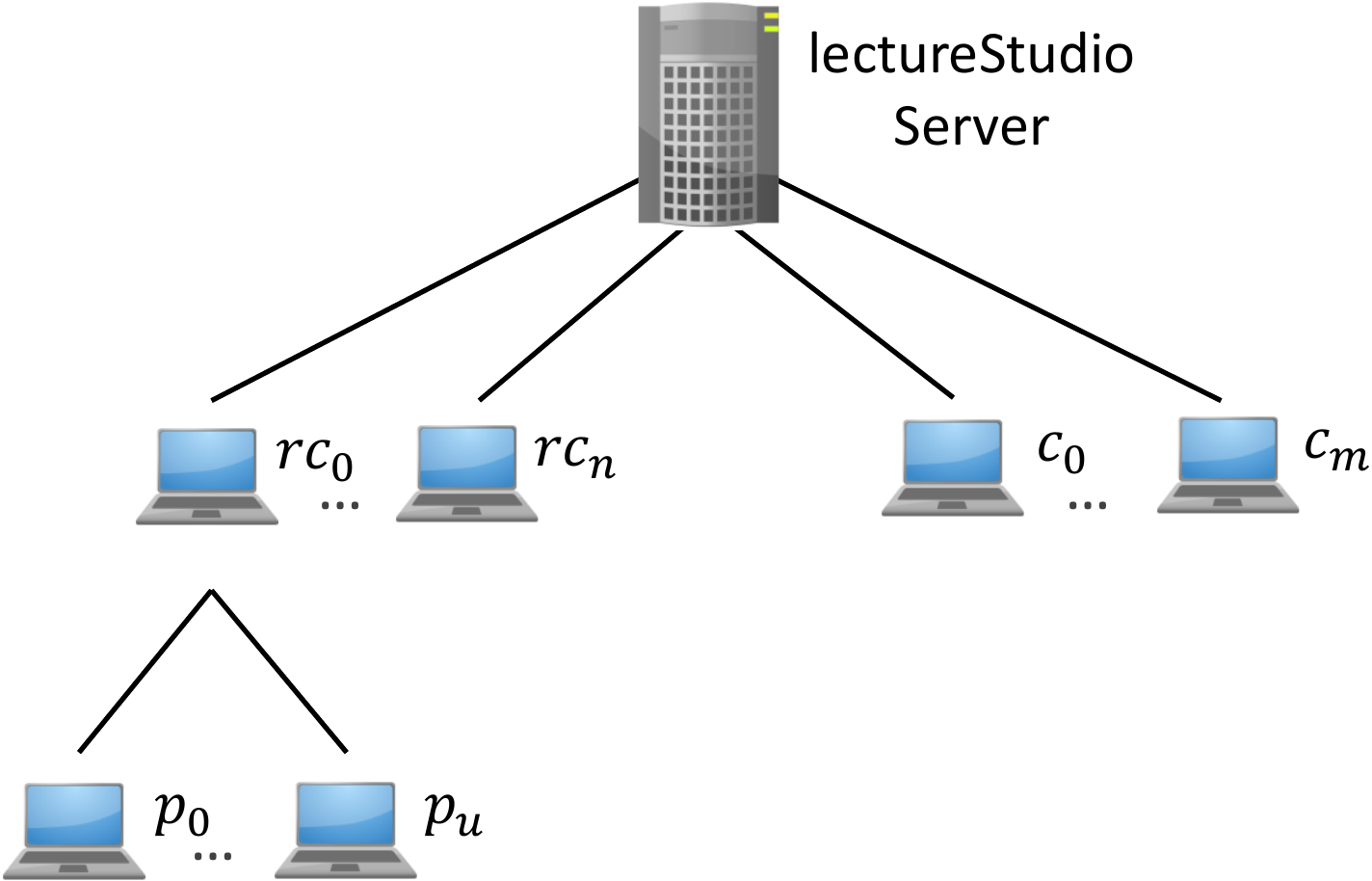}
    \caption{Example network with the following types of nodes: $c_0$ to $c_m$ are normal clients directly connected to the lectureStudio server, $rc_0$ to $rc_n$ are relay clients, and $p_0$ to $p_u$ are relayed clients, i.e., clients that are not directly connected to the lectureStudio server.}
    \label{figure:ilp-constraints}
\end{figure}

As an example of the necessary global constraints used in our lectureStudio scenario, the following paragraph explains three important (simplified) \ac{ILP} constraints that model inter-rule conditions.
This section distinguishes three types of client nodes as shown in \cref{figure:ilp-constraints}:
$c_0$ to $c_m$ denote normal clients directly connected to the lectureStudio server, $rc_0$ to $rc_n$ are relay clients, and $p_0$ to $p_u$ represent relayed clients, i.e., clients that are not directly connected to the lectureStudio server.

In our approach, every new client with the index $i$ must either be directly connected to the lectureStudio server ($c_i$), be converted to a relay client ($rc_i$), or be connected to another node that is a relay client ($p_i$).
The corresponding binary variables ($c_i, rc_i,$ and $p_i$) are set to $1$ if the respective case occurs.
The \ac{ILP} constraint given in \cref{eq-ilp-mapped-once} ensures the explained condition for every new client.
Note that $NC$ denotes the set of all new clients.
\begin{equation}
    \forall i \in \mathbb{N} \enspace \text{with} \enspace 0 \leq i < |NC| \enspace : \enspace rc_i + c_i + p_i = 1, \enspace rc_i, \,c_i, \,p_i \in \{0,1\}\label{eq-ilp-mapped-once}
\end{equation}
Furthermore, for every (new) relay client with the index $i$ there must be at least one newly added \ac{P2P} connection to other clients because a regular client should only be converted to a relay client if it actually relays the data to another client.
The variable $p2p_{ij}$ denotes a new \ac{P2P} connection that will be created via a respective \ac{GT} rule to connect node $i$ and node $j$.
The condition can be expressed with the \ac{ILP} constraint listed in \cref{eq-ilp-relay-client}.
\begin{equation}
    \forall i \in \mathbb{N} \enspace \text{with} \enspace 0 \leq i < |NC| \enspace : \enspace rc_i \leq \sum_{j=0}^{n} {p2p}_{ij}, \enspace rc_i, \,{p2p}_{ij} \in \{0,1\} \enspace \text{and} \enspace n \in \mathbb{N}^+\label{eq-ilp-relay-client}
\end{equation}
Additionally, it must be ensured that a \ac{P2P} connection does not use more bandwidth ($bw_{ij}$) than what is available on the internet link ($con_{ij}$) between two respective nodes with the indices $i$ and $j$.
This behavior can be achieved with the (simplified) \ac{ILP} constraint given in \cref{eq-ilp-bandwidth}, in which $P$ denotes the set of all clients connected to relay clients and $RC$ represents the set of all relay clients.
\begin{equation}
    \forall i,j \in \mathbb{N} \enspace \text{with} \enspace 0 \leq i < |P| \enspace \text{and} \enspace 0 \leq j < |RC| \enspace : \enspace x_{ij} \cdot {bw}_{ij} \leq x_{ij} \cdot {con}_{ij}, \enspace bw_{ij}, \,con_{ij} \in \mathbb{N}^+\label{eq-ilp-bandwidth}
\end{equation}
The complete (and non-simplified) \langname~definition with all the necessary constraints and the objective of the \textit{Plan} stage can be found on our GitHub repository\footnote{\langname~file for the lectureStudio scenario - \url{https://github.com/Echtzeitsysteme/gips-gcm-2023-example/blob/main/org.gips.examples.incrementalp2p.gips.incrementaldistribution/src/gipsl/Model.gipsl}}.

The \textit{Monitor} block must currently be implemented by hand and cannot be generated from a specification, despite our generic approach in \toolname.
This is due to the fact, that the \textit{Monitor} represents a bridge from the managed system to the knowledge base, which is a highly domain-specific problem and depends on the used technologies.
A \textit{Monitor} that is connected to a simulation differs greatly from a \textit{Monitor} that is connected to a server application via the internet.

After a full cycle of the MAPE-K loop, the program is ready to adapt the state of the overlay network again based on the changes that occurred in the meantime.
Hence, our example implementation allows for an incremental adaption of the \ac{P2P} overlay network as a reaction to model changes.

% Demonstration
\subsection{Prototype Demonstration}
\label{subsec:demonstration}

We have prepared an example of the calculation of the lectureStudio \ac{P2P} overlay network.
\Cref{figure:example-graph} shows screenshots of the resulting network visualizations of the example implementation.
The central entity (\texttt{Root-Server}) is connected to two relay clients (\texttt{Client3} and \texttt{Client11}).
All normal clients (blue) are either connected to the relay clients (violet) or to the \texttt{Root-Server} (pink).
The edge thickness denotes the available bandwidth between the nodes.
The thicker it is, the more bandwidth is available.

\newcommand{\demospacing}{0.49}

\begin{figure}[hbt]
\centering
    \begin{subfigure}[t]{\demospacing\textwidth}
        \centering
        \includegraphics[width=1.0\linewidth,clip,trim=6cm 3cm 15cm 5cm]{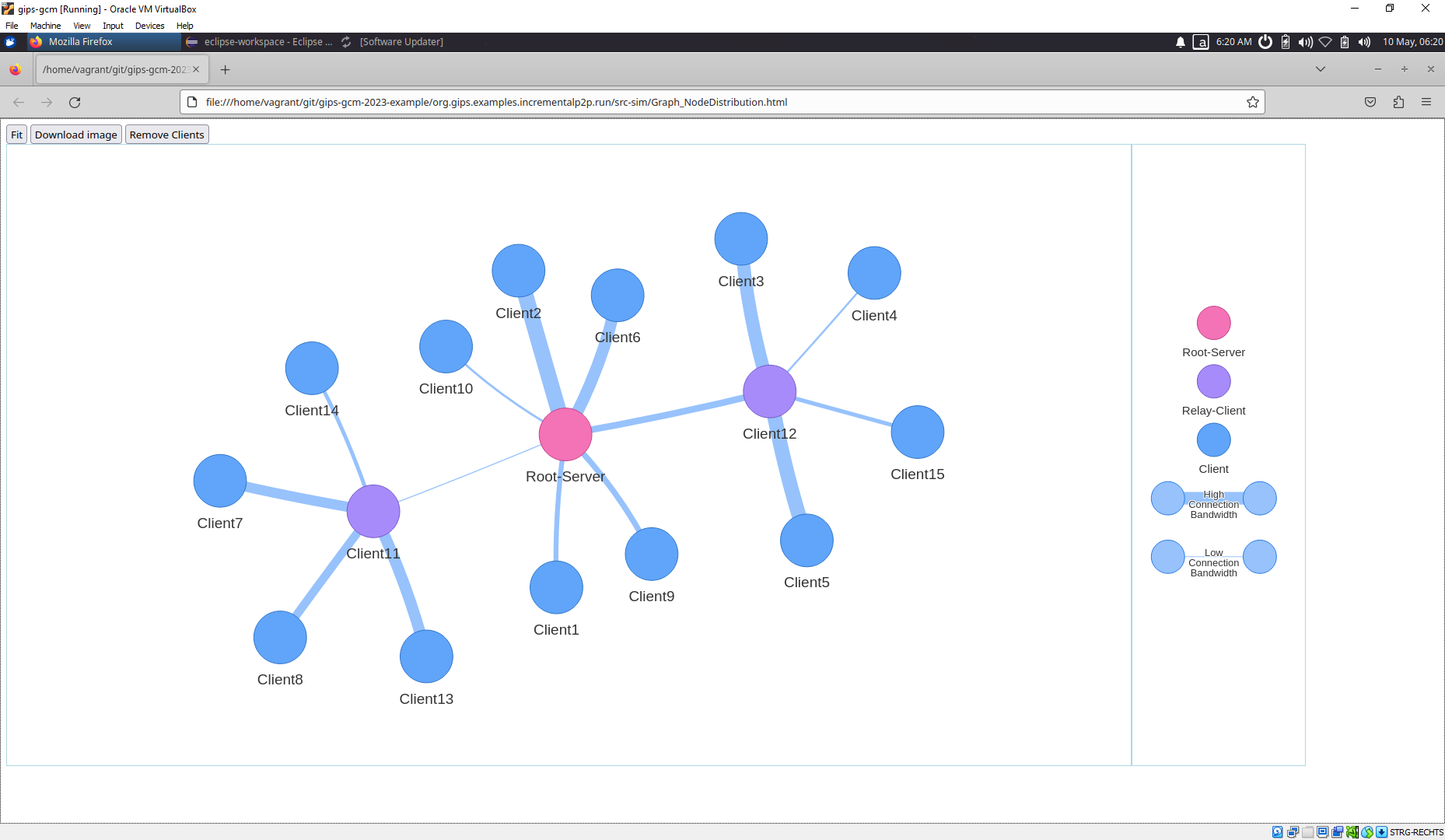}
        \caption{Initial network state.}
        \label{figure:example-graph-1}
    \end{subfigure}
    ~
    \begin{subfigure}[t]{\demospacing\textwidth}
        \centering
        \includegraphics[width=1.0\linewidth,clip,trim=6cm 3cm 15cm 5cm]{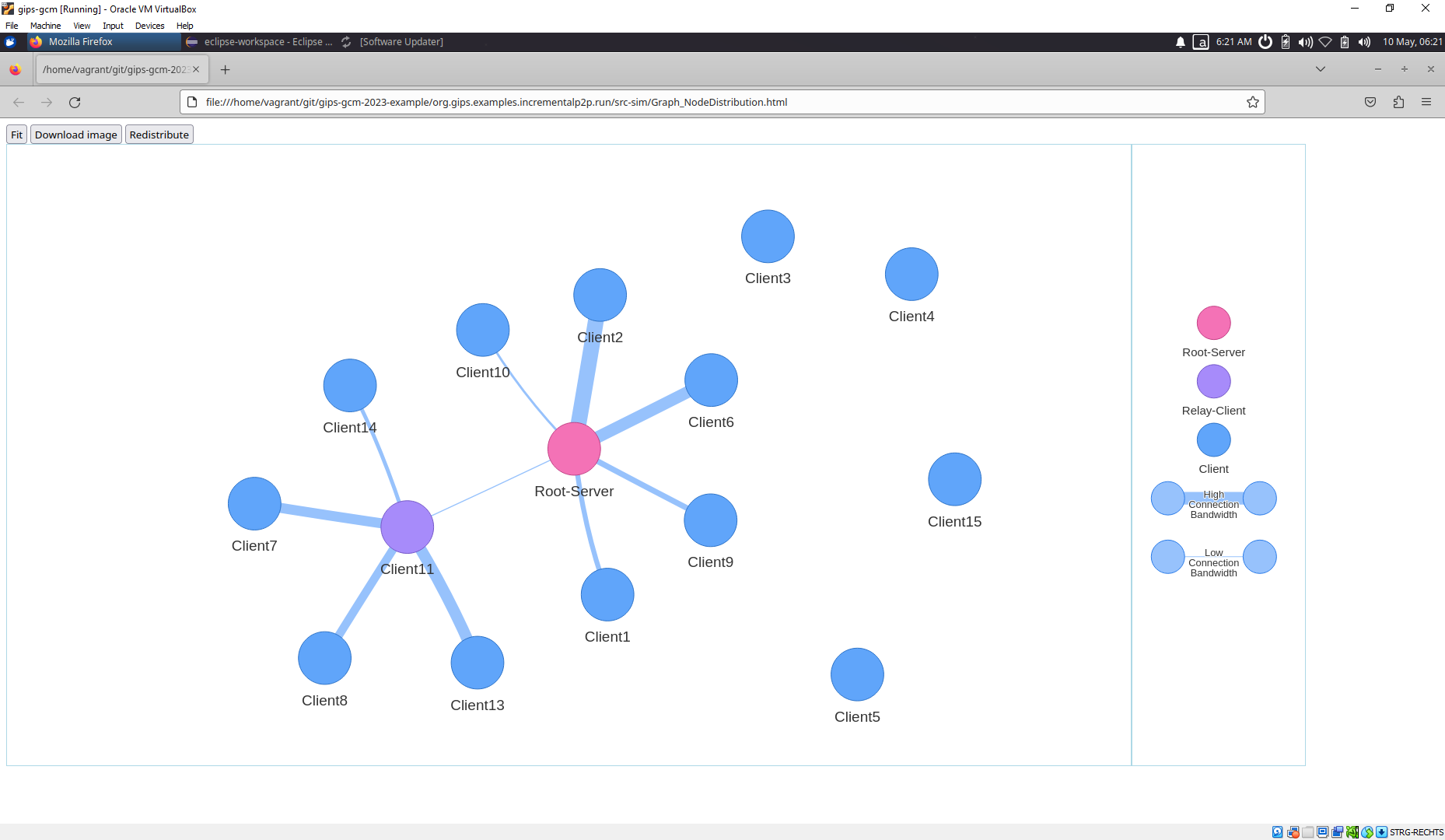}
        \caption{Second relay client was removed.}
        \label{figure:example-graph-2}
    \end{subfigure}
    ~
    \begin{subfigure}[t]{\demospacing\textwidth}
        \centering
        \includegraphics[width=1.0\linewidth,clip,trim=6cm 3cm 15cm 5cm]{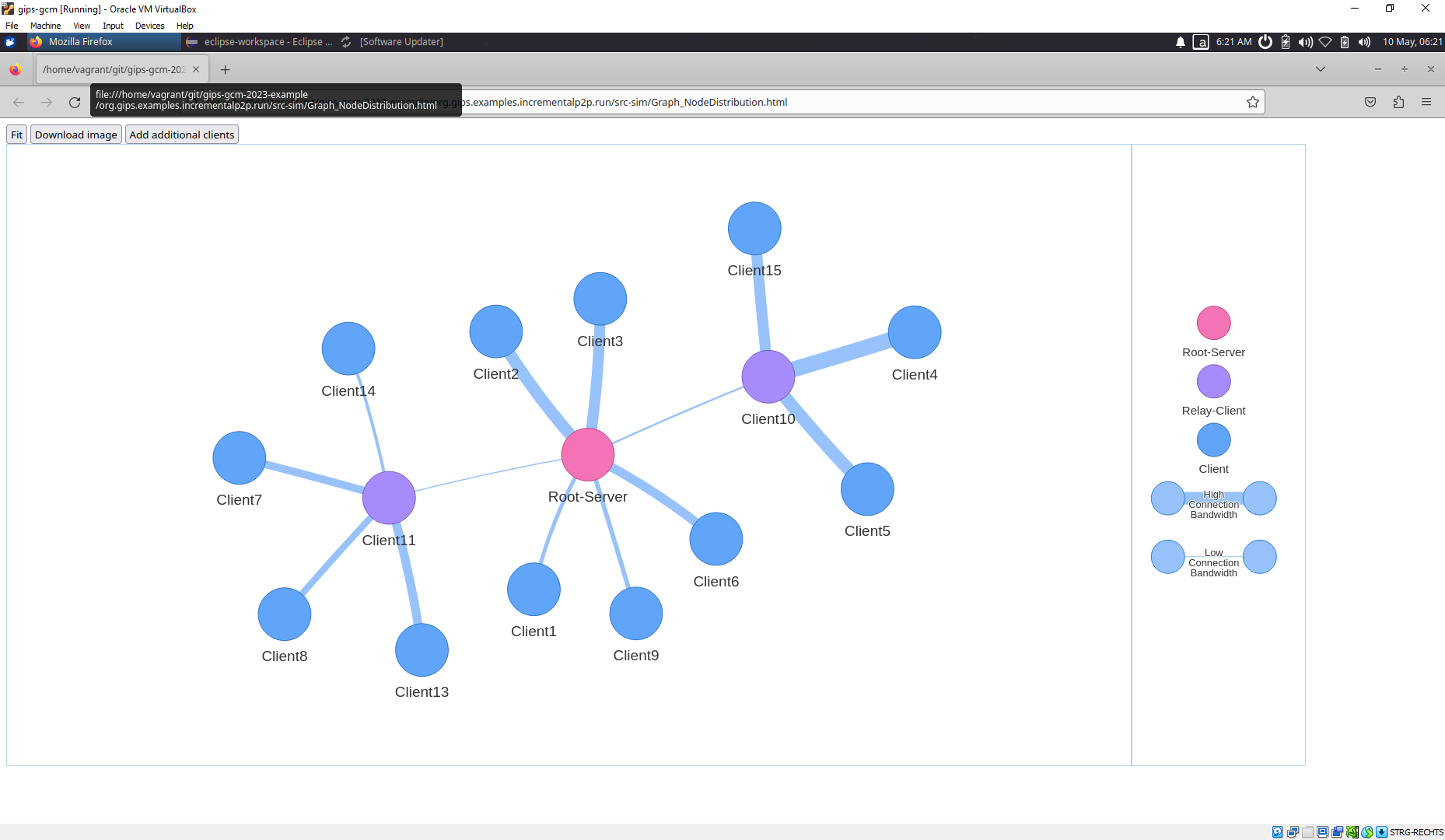}
        \caption{Floating clients were reconnected.}
        \label{figure:example-graph-3}
    \end{subfigure}
    ~
    \begin{subfigure}[t]{\demospacing\textwidth}
        \centering
        \includegraphics[width=1.0\linewidth,clip,trim=6cm 3cm 15cm 5cm]{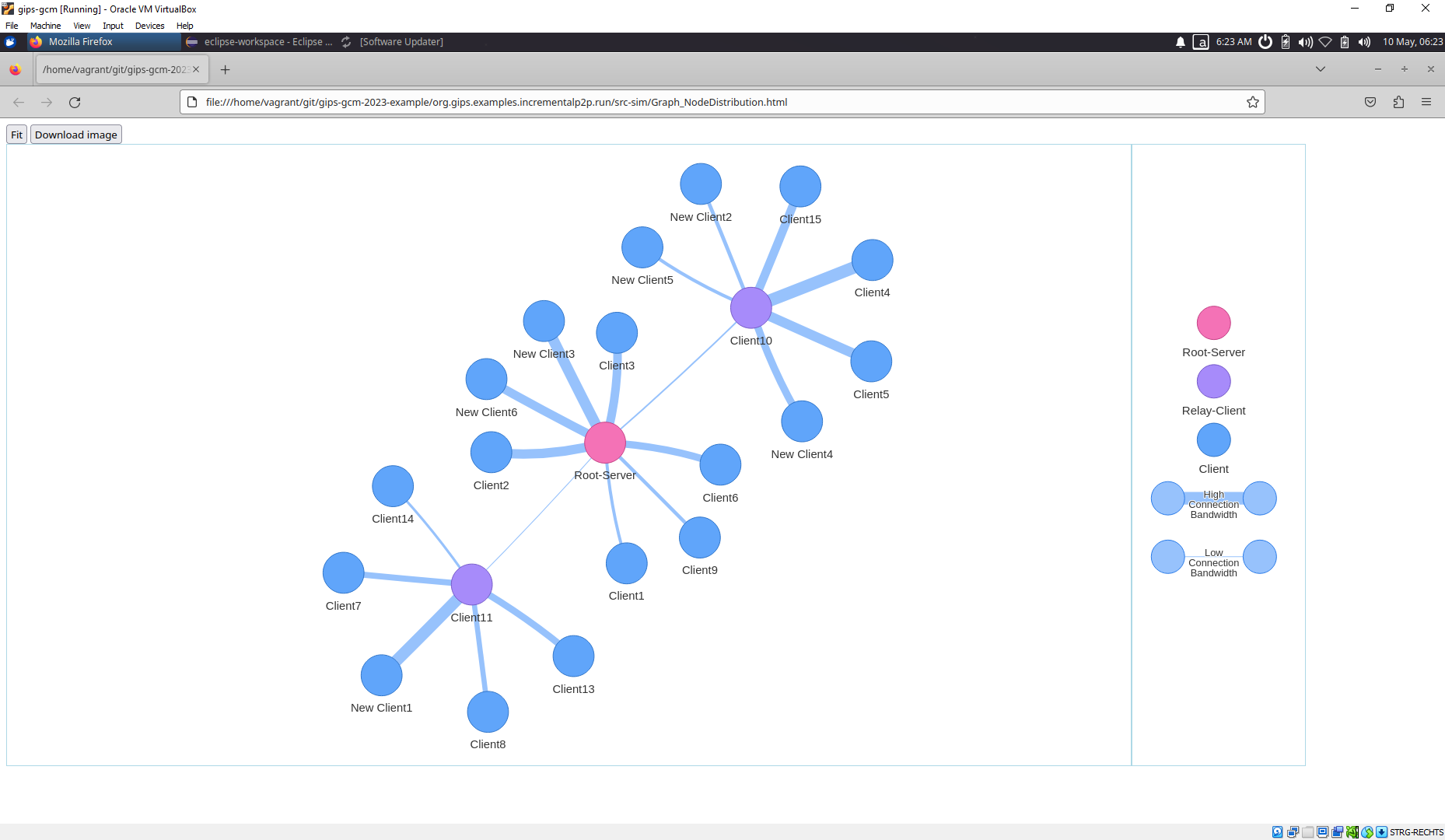}
        \caption{Additional clients were connected.}
        \label{figure:example-graph-4}
    \end{subfigure}
    \caption{Screenshots of the example's network visualizations.}
    \label{figure:example-graph}
\end{figure}

In this scenario, there will be one central \texttt{LectureStudioServer} instance and $15$\,clients in the initial setup (see \cref{figure:example-graph-1}).
As a next step, one of the relay clients will be removed from the model (see \cref{figure:example-graph-2}), e.g., because the corresponding student left the lecture stream.
The incremental implementation is able to formulate the necessary network repair operation as an \ac{ILP} problem\footnote{The complete \langname~definition for all \ac{ILP} constraints can be found on GitHub - \url{https://github.com/Echtzeitsysteme/gips-gcm-2023-example/blob/main/org.gips.examples.incrementalp2p.gips.incrementaldistribution/src/gipsl/Model.gipsl}} as well and calculates a set of \ac{GT} rule applications to correct the network, i.e., to re-connect the clients that lost their connection.
The result is shown in \cref{figure:example-graph-3}.
To further show the capabilities of the example implementation, the next step adds $6$\,new clients, that represent, for example, new students joining the stream.
The implementation will automatically add these new clients to the data distribution overlay network, which is shown in \cref{figure:example-graph-4}.
The result can be reproduced with the artifact that consists of the \toolname~framework and the presented implementation.
To ease the installation for interested readers, we created a virtual machine\footnote{GIPS GCM 2023 Artifact VM - \url{https://github.com/Echtzeitsysteme/gips-gcm-2023-artifact-vm}} with all required software installed and documentation to be able to modify, execute, and visualize the example.
We used the open-source solver GLPK because it does not need any manual licensing.
A typical execution of our example finishes after an interval of \SI{2}{\second} to \SI{30}{\second} if run inside the virtual machine.
The result of a simulation run will be displayed in the web browser within the virtual machine.

\begin{figure*}[hbt]
\centering
    \includegraphics[width=0.7\linewidth]{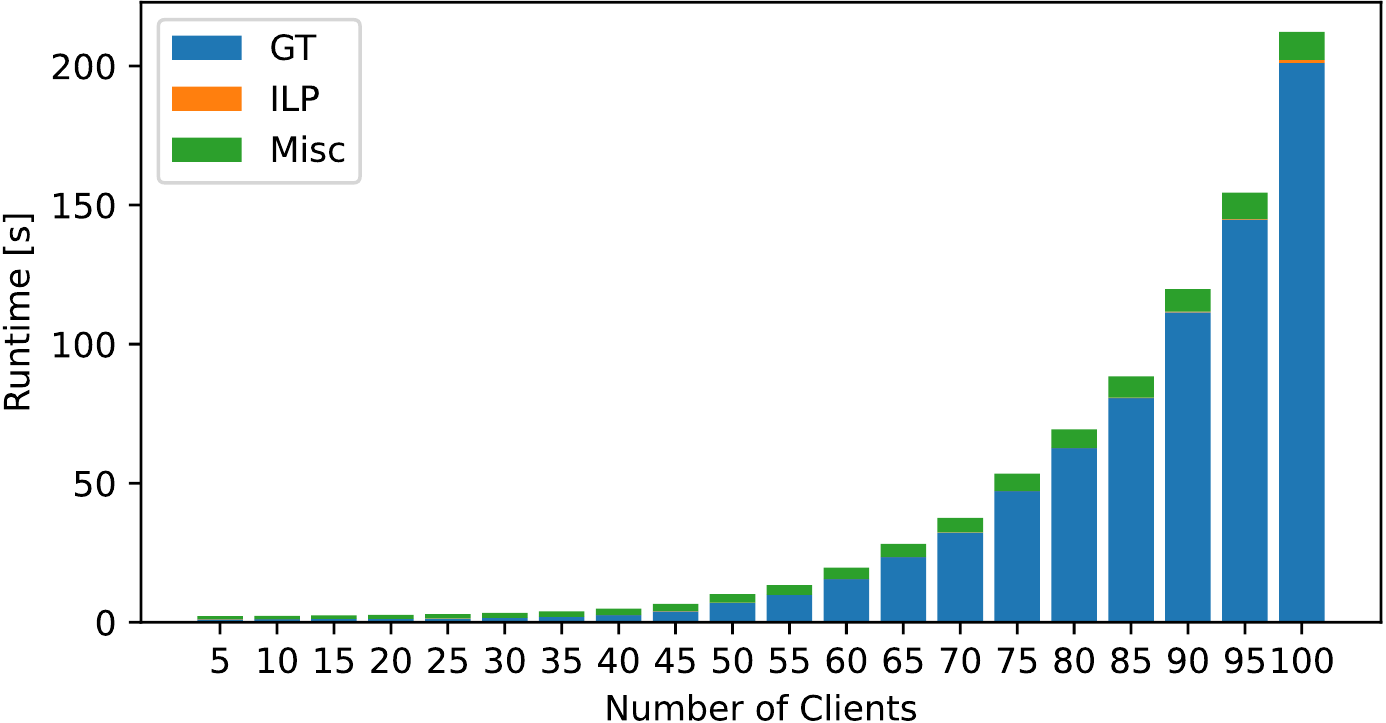}
    \caption{Aggregated algorithm runtime (y-axis) for an individual number of clients (x-axis) incrementally joining a lecture. The individual runtimes are split into \ac{GT}, \ac{ILP}, and miscellaneous parts.}
    \label{figure:eval-plot}
\end{figure*}

% Evaluation
Of course, the question of how well the presented approach scales w.r.t.\@ the number of clients within the streaming network arises.
To answer this question, we ran experiments with different numbers of clients that incrementally join the stream.
In this scenario, our generated topology control algorithm connects all newly arriving clients to the \ac{P2P} network, while minimizing the overall document distribution time.
In the evaluation setup, the \texttt{LectureStudioServer} is connected to the network via a \SI{150}{\mega\bit\per\second} connection and no initial \texttt{Client} has joined, yet.
The upload and download bandwidths of all \texttt{Client}s are sampled from Speedtest.net measurements\footnote{Speedtest.net median bandwidth in Germany - \url{https://www.speedtest.net/global-index/germany}}.
Every result shown in this section is the calculated mean of ten measurements.
All experiments were run on a workstation equipped with an AMD Ryzen Threadripper 2990WX with 32\,CPU cores and \SI{128}{\giga\byte} of memory.
The operating system used is Ubuntu 20.04.6 LTS, the Java environment used is OpenJDK Temurin (build 17.0.2+8), and the ILP solver is GLPK.
\Cref{figure:eval-plot} shows runtimes (y-axis) of the algorithm for each individual number of clients (x-axis), additionally, separated into \ac{GT}, \ac{ILP}, and miscellaneous parts.
The results show that the generated algorithm calculates a \ac{P2P} topology for up to $75$ clients in under one minute.
Unfortunately, \cref{figure:eval-plot} shows a large growth of the runtime when scaling up the number of \texttt{Client}s.
Interestingly, the majority of the runtime needed to calculate each result is caused by the pattern-matching process of the \ac{GT} engine.
Thus, the time needed by GLPK to solve each resulting \ac{ILP} problem is barely visible.
The exponential increase in runtime is most likely a result of the used graph patterns.
Some of the patterns contain multiple disjoint pattern nodes, which results in match sets that grow exponentially with an increasing number of clients.

When using \toolname~in practice, sometimes the \ac{GT} part and sometimes the \ac{ILP} part dominate the runtime behavior.
The sum of both is often lower than the runtime of a purely \ac{ILP}-based approach.
Corresponding comparisons of naive generation of the \ac{ILP} problem with \ac{GT} rules with very low \ac{GT} runtime and almost the entire runtime of the \ac{ILP} solver, as well as a heavier preprocessing by \ac{GT} rules, are planned for a future publication.

\subsection{Challenges in leaving the Prototype Phase}
\label{subsec:threats-to-validity}

In the previous sections, we have shown how to use \toolname~as a tool for prototyping a control algorithm using an example of the topology control domain. 
More specifically, we have used \toolname~to automatically derive MAPE-K building blocks from a \langname~specification.
From a rapid prototyping point of view, one can see that \toolname~is an excellent tool to facilitate the development of (topology) control algorithms, since \toolname~can be used to develop and evaluate prototypes by experimenting with parameters, different objective functions, and different constraints.

A possible design challenge is the fact that the result of a \toolname~calculation is a set of \ac{GT} rule matches without any additional information about a suitable execution order that would guarantee that all selected matches are processed.
To guarantee that a \ac{GT} rule application does not invalidate a match of another \ac{GT} rule, developers must introduce additional \langname~constraints, if this is a necessary requirement for the specific problem domain.

While \toolname~is a promising approach for developing prototypes, it is clear that a control algorithm based on \ac{ILP} and \ac{GT} will not be suitable for most real-world application scenarios that have to work under real-time constraints (e.g., streaming, content distribution, etc.).
However, the developed prototypes can be used to calculate solutions for offline optimization problems, e.g., static task scheduling.
Furthermore, there are other application domains, for example, the placement of virtual networks in data centers, for which \toolname~with its \ac{ILP} and \ac{GT}-based algorithms can compete with state-of-the-art approaches \cite{gipsGCM}.
In such a problem domain, the generated algorithms scale reasonably well and, moreover, the developers are able to try and evaluate different heuristic approaches, too.
Once a \toolname-based prototype is developed far enough to satisfy the desired specifications, one can use the results of the prototype to create an optimized implementation using, e.g., the C++ programming language for further acceleration.
In the \ac{ML} community the approach of programmatic labeling (programmatic weak supervision) is well known, where data sets are automatically annotated for training \ac{ML} using exact algorithms, heuristics, or other \ac{ML} approaches \cite{DBLP:journals/corr/abs-2202-05433}.
We propose to implement an exact algorithm using our \toolname~framework to annotate data sets for an \ac{ML} approach, e.g., neural networks, automatically.
This means that we take automatically generated problem instances and solve them using the \toolname~algorithm to convert the obtained results to a training data set.
In fact, this is an approach we are currently pursuing and will present in a future paper.

\section{Related Work}
\label{sec:related-work}

We have identified a general need for model-driven rapid prototyping because of the ever-growing complexity of modern software systems.
To be able to relate \toolname~to the current state-of-the-art, we will list a few related works of the self-adaptive systems community as well as selected works of the software engineering community, which touch on the same subjects as our work such as, e.g., graph transformation, optimization, MAPE-K loop development, and resource allocation.

% Becker/Giese et al.
Becker \etal\cite{giese} defined the requirements to model correct self-adaptive systems on a high level of abstraction and presented a \ac{GT}-based formal solution to most of the identified requirements.
The authors explain how UML class and object diagrams together with the behavioral modeling of \ac{GT} can be used to model correct self-adaptive systems.
This concept is evaluated using a simplified application example with a network of pipes and filters.
Similar to our approach in \toolname, Becker \etal used \ac{GT} to express local structural constraints and model transitions, while \toolname, on the other hand, is also able to express and enforce global constraints on the model by using \ac{ILP}.

% SimSOTA
Abeywickrama \etal\cite{simsota} presented a novel approach to engineer interacting, centralized, and decentralized feedback loops using the state of the affairs model \cite{sota} to express self-awareness and self-adaption.
Therefore, their notion of feedback loops extends the well-known MAPE-K adaption model \cite{visionOfAutonomicComputing, Brun2009}. 
In their work, they implemented an Eclipse plug-in for the modeling, simulation, and validation of self-adaptive systems based on their feedback-loop-based approach.
Compared to \toolname, their Eclipse plug-in is currently unable to derive runtime code (e.g., Java) from the models.

% MOMoT
Fleck \etal\cite{fleck2016SearchBasedModelTransformations} presented a model-driven approach to tackle the rule orchestration problem.
They try to find an ordered set of rule-match pairs to modify a model to optimize a given fitness function.
In contrast to \toolname, the result is a multi-objective optimization problem intended to be solved by a search-based optimization algorithm, for example, a genetic algorithm.
A downside of this approach is the fact that genetic algorithms cannot guarantee optimal solutions and, furthermore, can have problems finding promising sequences of rule applications in the large search space.
If runtime is not an issue, \toolname~can guarantee optimal solutions by utilizing the \ac{ILP} solver.
In contrast to \toolname, the approach by Fleck \etal does not need annotated rules or the specification of additional constraints to find the rule application sequence.

% MDEOptimiser
With the framework MDEOptimiser Burdusel \etal\cite{mdeoptimiser} present another search-based solution to find sequences of rule applications that, when applied, lead to an optimal model state. 
In contrast to \cite{fleck2016SearchBasedModelTransformations}, which performs rule-based optimization, Burdusel \etal follow the model-based optimization approach. 
In essence, they store each evolution of the initial model as a self-contained new model during the optimization process, instead of only storing the changes (i.e., rule applications) between each optimization step. 
Similar to \cite{fleck2016SearchBasedModelTransformations}, Burdusel \etal cannot guarantee optimal solutions, if they exist, due to the usage of evolutionary algorithms as means of optimization. 

% John, Taentzer
John \etal\cite{john2023} make use of MDEOptimiser to demonstrate and practically evaluate their new formal framework for model-based optimizations, in which they define completeness and soundness as new criteria for mutation operators.

% Chen - MAPE-K loop + SMT + QVT(-R)
Chen \etal\cite{10.1145/2568225.2568310} implement an incremental generative self-adaptation scheme based on model transformations at runtime. 
Similar to our approach, the authors implement a MAPE-K loop, where a planning component produces a set of \ac{GT} rule applications that, when applied, improve the utility of a given model.
In contrast to our approach, Chen \etal make use of \ac{SMT} to define their repair problem, use \ac{SMT} solvers to solve said problem, and implement their model transformations using medini QVT.
The former design choice makes it very hard to define a problem that, when solved, results in a solution that is not only valid (satisfies all constraints) but also optimal according to some objective function.
The latter design choice renders the tool unusable today since medini QVT was deprecated a while ago.

% Song - DSL, SMT
Song \etal\cite{songhui} do not present a self-adaption approach per se but, instead, present a framework similar to \toolname, with which one can implement algorithms that apply changes to a given model such that it satisfies a set of constraints.
Similar to Chen \etal\cite{10.1145/2568225.2568310}, Song \etal make use of \ac{SMT} to encode model constraints.
Furthermore, their approach comes with a very lean \ac{DSL}, mostly for configuration purposes, where model constraints can be defined using OCL expressions.
In contrast to \toolname, their approach does not give the means to define any model transformations with which a model could be changed, in order to satisfy a set of constraints.

% Giese/Ghahremani - MAPE-K loop + search-based optimization + 
What Ghahremani \etal\cite{10.1145/3380965} describe as utility-driven self-healing for dynamic architectures is, essentially, an approach for incremental model adaptation at runtime, which is in a way similar to Chen \etal\cite{10.1145/2568225.2568310}.
In contrast to Chen \etal and our approach, Ghahremani \etal do not specify an \ac{SMT} nor an \ac{ILP} problem; instead, they implement a search-based approach that aims to improve the result value of a given utility function.
Said utility function is composed of several (possibly user-defined) sub-utility functions that each evaluate matches belonging to the \ac{LHS} of a corresponding "repair" rule.
The goal is to find a set of rule applications that maximize the utility function and improve or "heal" the given model.

\section{Conclusion and Future Work}
\label{sec:conclusion-future-work}

% Conclusion
This paper presented an approach for the rapid prototype development of control algorithms, aided by \ac{GT}, \ac{ILP}, and code generation from a high-level specification.
Therefore, we briefly introduced the \toolname~framework and explained an example algorithm for the maintenance of \ac{P2P} overlay networks, which was, for the most part, generated by \toolname~from a high-level specification, in an effort to demonstrate \toolname's capabilities as a rapid prototyping software development tool.
To provide these capabilities, our framework was designed with \ac{MDSE} principles in mind and, thus, generates software artifacts from a given \langname~specification.
With our example of the \ac{P2P} overlay network maintenance tool for lectureStudio, we demonstrated the automatic derivation of building blocks for control algorithms to support the development of MAPE-K loops.
By means of \ac{ILP}-based optimization our algorithm sets up and maintains an overlay network for the streaming platform lectureStudio with the goal to minimize the required bandwidth for file transfers while upholding an acceptable document distribution time for all participants.
From our point of view, this example demonstrates the usefulness of the \toolname~framework for rapid prototyping to facilitate research within the domain of (topology) control algorithms.

% Future work
In the future, we want to gather more information on how \toolname~can be used in research of various (possibly more complex) domains, by implementing and evaluating a variety of other scenarios, e.g., the static scheduling of tasks on CPU processors.
Regarding the control algorithm domain, it is still an open question how or if the \textit{Monitor} block of the MAPE-K loop can be generated automatically.
Moreover, it would be interesting to include a mechanism to support the generation of valid \ac{GT} rule application sequences (instead of sets) to guarantee the validity of the \toolname~output.
Finally, as an ongoing effort, we plan to extend the expressiveness of \langname~further, by developing new language features and shortcuts to ease the specification process for tool designers of various domains.

% Artifact download
The presented example containing the \toolname~framework, an open-source \ac{ILP} solver, the algorithm's source code, and its documentation, is publicly available on GitHub\footnote{GIPS GCM 2023 Artifact VM - \url{https://github.com/Echtzeitsysteme/gips-gcm-2023-artifact-vm}} as a virtual machine.

%
% Acknowledgment(s)
%
\paragraph{Acknowledgements}
This work has been funded by the Deutsche Forschungsgemeinschaft (DFG, German Research Foundation) - Project-ID 210487104 - SFB 1053.

\nocite{*}
\bibliographystyle{eptcs}
\bibliography{literature}

\begin{thebibliography}{10}
\providecommand{\bibitemdeclare}[2]{}
\providecommand{\surnamestart}{}
\providecommand{\surnameend}{}
\providecommand{\urlprefix}{Available at }
\providecommand{\url}[1]{\texttt{#1}}
\providecommand{\href}[2]{\texttt{#2}}
\providecommand{\urlalt}[2]{\href{#1}{#2}}
\providecommand{\doi}[1]{doi:\urlalt{https://doi.org/#1}{#1}}
\providecommand{\eprint}[1]{arXiv:\urlalt{https://arxiv.org/abs/#1}{#1}}
\providecommand{\bibinfo}[2]{#2}

\bibitemdeclare{inproceedings}{sota}
\bibitem{sota}
\bibinfo{author}{Dhaminda~B. \surnamestart Abeywickrama\surnameend},
  \bibinfo{author}{Nicola \surnamestart Bicocchi\surnameend} \&
  \bibinfo{author}{Franco \surnamestart Zambonelli\surnameend}
  (\bibinfo{year}{2012}): \emph{\bibinfo{title}{SOTA: Towards a General Model
  for Self-Adaptive Systems}}.
\newblock In: {\slshape \bibinfo{booktitle}{2012 IEEE 21st Int. Workshop on
  Enabling Technologies: Infrastructure for Collaborative Enterprises}}, pp.
  \bibinfo{pages}{48--53}, \doi{10.1109/WETICE.2012.48}.

\bibitemdeclare{inproceedings}{simsota}
\bibitem{simsota}
\bibinfo{author}{Dhaminda~B. \surnamestart Abeywickrama\surnameend},
  \bibinfo{author}{Nicklas \surnamestart Hoch\surnameend} \&
  \bibinfo{author}{Franco \surnamestart Zambonelli\surnameend}
  (\bibinfo{year}{2013}): \emph{\bibinfo{title}{SimSOTA: Engineering and
  Simulating Feedback Loops for Self-Adaptive Systems}}.
\newblock In: {\slshape \bibinfo{booktitle}{Proc. of the Int. C* Conf. on
  Computer Science and Software Engineering}}, \bibinfo{series}{C3S2E '13},
  \bibinfo{publisher}{Association for Computing Machinery}, p.
  \bibinfo{pages}{67–76}, \doi{10.1145/2494444.2494446}.

\bibitemdeclare{book}{bazaraa2011linear}
\bibitem{bazaraa2011linear}
\bibinfo{author}{Mokhtar~S. \surnamestart Bazaraa\surnameend},
  \bibinfo{author}{John~J. \surnamestart Jarvis\surnameend} \&
  \bibinfo{author}{Hanif~D. \surnamestart Sherali\surnameend}
  (\bibinfo{year}{2011}): \emph{\bibinfo{title}{Linear programming and network
  flows}}.
\newblock \bibinfo{publisher}{John Wiley \& Sons}, \doi{10.1002/9780471703778}.

\bibitemdeclare{inproceedings}{giese}
\bibitem{giese}
\bibinfo{author}{Basil \surnamestart Becker\surnameend} \&
  \bibinfo{author}{Holger \surnamestart Giese\surnameend}
  (\bibinfo{year}{2008}): \emph{\bibinfo{title}{Modeling of Correct
  Self-Adaptive Systems: A Graph Transformation System Based Approach}}.
\newblock In: {\slshape \bibinfo{booktitle}{Proc. of the 5th Int. Conf. on Soft
  Computing as Transdisciplinary Science and Technology}},
  \bibinfo{series}{CSTST '08}, \bibinfo{publisher}{Association for Computing
  Machinery}, p. \bibinfo{pages}{508–516}, \doi{10.1145/1456223.1456326}.

\bibitemdeclare{article}{Bencomo2019}
\bibitem{Bencomo2019}
\bibinfo{author}{Nelly \surnamestart Bencomo\surnameend},
  \bibinfo{author}{Sebastian \surnamestart G{\"o}tz\surnameend} \&
  \bibinfo{author}{Hui \surnamestart Song\surnameend} (\bibinfo{year}{2019}):
  \emph{\bibinfo{title}{Models@run.time: a guided tour of the state of the art
  and research challenges}}.
\newblock {\slshape \bibinfo{journal}{Software {\&} Systems Modeling}}
  \bibinfo{volume}{18}(\bibinfo{number}{5}), pp. \bibinfo{pages}{3049--3082},
  \doi{10.1007/s10270-018-00712-x}.

\bibitemdeclare{article}{modelsAtRuntime}
\bibitem{modelsAtRuntime}
\bibinfo{author}{Gordon \surnamestart Blair\surnameend}, \bibinfo{author}{Nelly
  \surnamestart Bencomo\surnameend} \& \bibinfo{author}{Robert~B. \surnamestart
  France\surnameend} (\bibinfo{year}{2009}): \emph{\bibinfo{title}{Models@
  run.time}}.
\newblock {\slshape \bibinfo{journal}{Computer}}
  \bibinfo{volume}{42}(\bibinfo{number}{10}), pp. \bibinfo{pages}{22--27},
  \doi{10.1109/MC.2009.326}.

\bibitemdeclare{book}{appliedMathematicalProgramming}
\bibitem{appliedMathematicalProgramming}
\bibinfo{author}{Stephen~P. \surnamestart Bradley\surnameend},
  \bibinfo{author}{Arnoldo~C. \surnamestart Hax\surnameend} \&
  \bibinfo{author}{Thomas~L. \surnamestart Magnanti\surnameend}
  (\bibinfo{year}{1977}): \emph{\bibinfo{title}{Applied Mathematical
  Programming}}.
\newblock \bibinfo{publisher}{Addison-Wesley}.

\bibitemdeclare{inbook}{Brun2009}
\bibitem{Brun2009}
\bibinfo{author}{Yuriy \surnamestart Brun\surnameend},
  \bibinfo{author}{Giovanna \surnamestart Di~Marzo~Serugendo\surnameend},
  \bibinfo{author}{Cristina \surnamestart Gacek\surnameend},
  \bibinfo{author}{Holger \surnamestart Giese\surnameend},
  \bibinfo{author}{Holger \surnamestart Kienle\surnameend},
  \bibinfo{author}{Marin \surnamestart Litoiu\surnameend},
  \bibinfo{author}{Hausi \surnamestart M{\"u}ller\surnameend},
  \bibinfo{author}{Mauro \surnamestart Pezz{\`e}\surnameend} \&
  \bibinfo{author}{Mary \surnamestart Shaw\surnameend} (\bibinfo{year}{2009}):
  \emph{\bibinfo{title}{Engineering Self-Adaptive Systems through Feedback
  Loops}}, pp. \bibinfo{pages}{48--70}.
\newblock \bibinfo{publisher}{Springer}, \doi{10.1007/978-3-642-02161-9_3}.

\bibitemdeclare{inproceedings}{mdeoptimiser}
\bibitem{mdeoptimiser}
\bibinfo{author}{Alexandru \surnamestart Burdusel\surnameend},
  \bibinfo{author}{Steffen \surnamestart Zschaler\surnameend} \&
  \bibinfo{author}{Daniel \surnamestart Str\"{u}ber\surnameend}
  (\bibinfo{year}{2018}): \emph{\bibinfo{title}{MDEoptimiser: A Search Based
  Model Engineering Tool}}.
\newblock In: {\slshape \bibinfo{booktitle}{Proc. of the 21st ACM/IEEE Int.
  Conf. on Model Driven Engineering Languages and Systems: Companion
  Proceedings}}, \bibinfo{series}{MODELS '18}, \bibinfo{publisher}{Association
  for Computing Machinery}, p. \bibinfo{pages}{12–16},
  \doi{10.1145/3270112.3270130}.

\bibitemdeclare{inproceedings}{10.1145/2568225.2568310}
\bibitem{10.1145/2568225.2568310}
\bibinfo{author}{Bihuan \surnamestart Chen\surnameend}, \bibinfo{author}{Xin
  \surnamestart Peng\surnameend}, \bibinfo{author}{Yijun \surnamestart
  Yu\surnameend}, \bibinfo{author}{Bashar \surnamestart Nuseibeh\surnameend} \&
  \bibinfo{author}{Wenyun \surnamestart Zhao\surnameend}
  (\bibinfo{year}{2014}): \emph{\bibinfo{title}{Self-Adaptation through
  Incremental Generative Model Transformations at Runtime}}.
\newblock In: {\slshape \bibinfo{booktitle}{Proc. of the 36th Int. Conf. on
  Software Engineering}}, \bibinfo{series}{ICSE 2014},
  \bibinfo{publisher}{Association for Computing Machinery}, p.
  \bibinfo{pages}{676–687}, \doi{10.1145/2568225.2568310}.

\bibitemdeclare{inbook}{Cheng2014}
\bibitem{Cheng2014}
\bibinfo{author}{Betty H.~C. \surnamestart Cheng\surnameend},
  \bibinfo{author}{Kerstin~I. \surnamestart Eder\surnameend},
  \bibinfo{author}{Martin \surnamestart Gogolla\surnameend},
  \bibinfo{author}{Lars \surnamestart Grunske\surnameend},
  \bibinfo{author}{Marin \surnamestart Litoiu\surnameend},
  \bibinfo{author}{Hausi~A. \surnamestart M{\"u}ller\surnameend},
  \bibinfo{author}{Patrizio \surnamestart Pelliccione\surnameend},
  \bibinfo{author}{Anna \surnamestart Perini\surnameend},
  \bibinfo{author}{Nauman~A. \surnamestart Qureshi\surnameend},
  \bibinfo{author}{Bernhard \surnamestart Rumpe\surnameend},
  \bibinfo{author}{Daniel \surnamestart Schneider\surnameend},
  \bibinfo{author}{Frank \surnamestart Trollmann\surnameend} \&
  \bibinfo{author}{Norha~M. \surnamestart Villegas\surnameend}
  (\bibinfo{year}{2014}): \emph{\bibinfo{title}{Using Models at Runtime to
  Address Assurance for Self-Adaptive Systems}}, pp. \bibinfo{pages}{101--136}.
\newblock \bibinfo{publisher}{Springer}, \doi{10.1007/978-3-319-08915-7_4}.

\bibitemdeclare{inproceedings}{gipsGCM}
\bibitem{gipsGCM}
\bibinfo{author}{Sebastian \surnamestart Ehmes\surnameend},
  \bibinfo{author}{Maximilian \surnamestart Kratz\surnameend} \&
  \bibinfo{author}{Andy \surnamestart Sch\"urr\surnameend}
  (\bibinfo{year}{2022}): \emph{\bibinfo{title}{Graph-Based Specification and
  Automated Construction of ILP Problems}}.
\newblock In: {\slshape \bibinfo{booktitle}{{\rm Proc. of the Thirteenth Int.
  Workshop on} Graph Computation Models, {\rm Nantes, France, 6th July 2022}}},
  {\slshape \bibinfo{series}{Electronic Proceedings in Theoretical Computer
  Science}} \bibinfo{volume}{374}, \bibinfo{publisher}{Open Publishing
  Association}, pp. \bibinfo{pages}{3--22}, \doi{10.4204/EPTCS.374.3}.

\bibitemdeclare{book}{taentzer}
\bibitem{taentzer}
\bibinfo{author}{Hartmut \surnamestart Ehrig\surnameend},
  \bibinfo{author}{Karsten \surnamestart Ehrig\surnameend},
  \bibinfo{author}{Ulrike \surnamestart Prange\surnameend} \&
  \bibinfo{author}{Gabriele \surnamestart Taentzer\surnameend}
  (\bibinfo{year}{2006}): \emph{\bibinfo{title}{Fundamentals of Algebraic Graph
  Transformation}}.
\newblock \bibinfo{publisher}{Springer}, \doi{10.1007/3-540-31188-2}.

\bibitemdeclare{article}{fleck2016SearchBasedModelTransformations}
\bibitem{fleck2016SearchBasedModelTransformations}
\bibinfo{author}{Martin \surnamestart Fleck\surnameend},
  \bibinfo{author}{Javier \surnamestart Troya\surnameend} \&
  \bibinfo{author}{Manuel \surnamestart Wimmer\surnameend}
  (\bibinfo{year}{2016}): \emph{\bibinfo{title}{Search-based model
  transformations}}.
\newblock {\slshape \bibinfo{journal}{Journal of Software: Evolution and
  Process}}, pp. \bibinfo{pages}{1081--1117}, \doi{10.1002/smr.1804}.

\bibitemdeclare{article}{PatternMatchReteNetwork}
\bibitem{PatternMatchReteNetwork}
\bibinfo{author}{Charles~L. \surnamestart Forgy\surnameend}
  (\bibinfo{year}{1982}): \emph{\bibinfo{title}{Rete: A Fast Algorithm for the
  Many Pattern/Many Object Pattern Match Problem}}.
\newblock {\slshape \bibinfo{journal}{Artificial Intelligence}}, p.
  \bibinfo{pages}{17–37}, \doi{10.1016/0004-3702(82)90020-0}.

\bibitemdeclare{article}{10.1145/3380965}
\bibitem{10.1145/3380965}
\bibinfo{author}{Sona \surnamestart Ghahremani\surnameend},
  \bibinfo{author}{Holger \surnamestart Giese\surnameend} \&
  \bibinfo{author}{Thomas \surnamestart Vogel\surnameend}
  (\bibinfo{year}{2020}): \emph{\bibinfo{title}{Improving Scalability and
  Reward of Utility-Driven Self-Healing for Large Dynamic Architectures}}.
\newblock {\slshape \bibinfo{journal}{ACM Trans. Auton. Adapt. Syst.}}
  \bibinfo{volume}{14}(\bibinfo{number}{3}), \doi{10.1145/3380965}.

\bibitemdeclare{article}{john2023}
\bibitem{john2023}
\bibinfo{author}{Stefan \surnamestart John\surnameend}, \bibinfo{author}{Jens
  \surnamestart Kosiol\surnameend}, \bibinfo{author}{Leen \surnamestart
  Lambers\surnameend} \& \bibinfo{author}{Gabriele \surnamestart
  Taentzer\surnameend} (\bibinfo{year}{2023}): \emph{\bibinfo{title}{A
  graph-based framework for model-driven optimization facilitating impact
  analysis of mutation operator properties}}.
\newblock {\slshape \bibinfo{journal}{Software and Systems Modeling}},
  \doi{10.1007/s10270-022-01078-x}.

\bibitemdeclare{article}{visionOfAutonomicComputing}
\bibitem{visionOfAutonomicComputing}
\bibinfo{author}{Jeffrey~O. \surnamestart Kephart\surnameend} \&
  \bibinfo{author}{David~M. \surnamestart Chess\surnameend}
  (\bibinfo{year}{2003}): \emph{\bibinfo{title}{The vision of autonomic
  computing}}.
\newblock {\slshape \bibinfo{journal}{Computer}}
  \bibinfo{volume}{36}(\bibinfo{number}{1}), pp. \bibinfo{pages}{41--50},
  \doi{10.1109/MC.2003.1160055}.

\bibitemdeclare{book}{luenbergerLinear2016}
\bibitem{luenbergerLinear2016}
\bibinfo{author}{David~G. \surnamestart Luenberger\surnameend} \&
  \bibinfo{author}{Yinyu \surnamestart Ye\surnameend} (\bibinfo{year}{1984}):
  \emph{\bibinfo{title}{Linear and Nonlinear Programming}}.
\newblock \bibinfo{publisher}{Springer}, \doi{10.1007/978-3-319-18842-3}.

\bibitemdeclare{inproceedings}{songhui}
\bibitem{songhui}
\bibinfo{author}{Hui \surnamestart Song\surnameend}, \bibinfo{author}{Xiaodong
  \surnamestart Zhang\surnameend}, \bibinfo{author}{Nicolas \surnamestart
  Ferry\surnameend}, \bibinfo{author}{Franck \surnamestart Chauvel\surnameend},
  \bibinfo{author}{Arnor \surnamestart Solberg\surnameend} \&
  \bibinfo{author}{Gang \surnamestart Huang\surnameend} (\bibinfo{year}{2014}):
  \emph{\bibinfo{title}{Modelling Adaptation Policies as Domain-Specific
  Constraints}}.
\newblock In: {\slshape \bibinfo{booktitle}{Model-Driven Engineering Languages
  and Systems}}, \bibinfo{publisher}{Springer}, pp. \bibinfo{pages}{269--285},
  \doi{10.1007/978-3-319-11653-2_17}.

\bibitemdeclare{article}{tomaszek2021VneEnsuringCorrectness}
\bibitem{tomaszek2021VneEnsuringCorrectness}
\bibinfo{author}{Stefan \surnamestart Tomaszek\surnameend},
  \bibinfo{author}{Roland \surnamestart Speith\surnameend} \&
  \bibinfo{author}{Andy \surnamestart Sch{\"u}rr\surnameend}
  (\bibinfo{year}{2021}): \emph{\bibinfo{title}{Virtual network embedding:
  ensuring correctness and optimality by construction using model
  transformation and integer linear programming techniques}}.
\newblock {\slshape \bibinfo{journal}{Software and Systems Modeling}}, pp.
  \bibinfo{pages}{1299--1332}, \doi{10.1007/s10270-020-00852-z}.

\bibitemdeclare{article}{DBLP:journals/corr/abs-2202-05433}
\bibitem{DBLP:journals/corr/abs-2202-05433}
\bibinfo{author}{Jieyu \surnamestart Zhang\surnameend},
  \bibinfo{author}{Cheng{-}Yu \surnamestart Hsieh\surnameend},
  \bibinfo{author}{Yue \surnamestart Yu\surnameend}, \bibinfo{author}{Chao
  \surnamestart Zhang\surnameend} \& \bibinfo{author}{Alexander \surnamestart
  Ratner\surnameend} (\bibinfo{year}{2022}): \emph{\bibinfo{title}{A Survey on
  Programmatic Weak Supervision}}.
\newblock {\slshape \bibinfo{journal}{CoRR}} \bibinfo{volume}{abs/2202.05433},
  \doi{10.48550/arXiv.2202.05433}.

\end{thebibliography}
\end{document}